\documentclass[showpacs,10pt,twocolumn,prb]{revtex4-1}

\usepackage{amsmath}
\usepackage{amssymb}
\usepackage{graphicx}
\usepackage{amssymb}
\usepackage{graphics}
\usepackage{epsfig}
\usepackage{CJK}
\usepackage{color}

\begin{document}

\title{Absence of Dirac states in BaZnBi$_{2}$ induced by spin-orbit coupling}
\author{Weijun Ren,$^{1,2}$ Aifeng Wang,$^{1}$ D. Graf,$^{3}$ Yu Liu,$^{1}$ Zhidong Zhang,$^{2}$ Wei-Guo Yin$^{1}$ and C. Petrovic$^{1}$}
\affiliation{$^{1}$Condensed Matter Physics and Materials Science Department, Brookhaven National Laboratory, Upton, New York 11973, USA\\
$^{2}$Shenyang National Laboratory for Materials Science, Institute of Metal Research, Chinese Academy of Sciences, Shenyang 110016, China\\
$^{3}$National High Magnetic Field Laboratory, Florida State University, Tallahassee, Florida 32306-4005, USA}

\date{\today}

\begin{abstract} We report magnetotransport properties of BaZnBi$_{2}$ single crystals. Whereas electronic structure features Dirac states, such states are removed from the Fermi level by spin-orbit coupling (SOC) and consequently electronic transport is dominated by the small hole and electron pockets. Our results are consistent with three dimensional (3D) but also with quasi two dimensional (2D) portions of the Fermi surface. The spin-orbit coupling-induced gap in Dirac states is much larger when compared to isostructural SrMnBi$_{2}$. This suggests that not only long range magnetic order but also mass of the alkaline earth atoms A in ABX$_{2}$ (A = alkaine earth, B = transition metal and X=Bi/Sb) are important for the presence of low-energy states obeying the relativistic Dirac equation at the Fermi surface.
\end{abstract}

\maketitle

\section{INTRODUCTION}

Rapid development of graphene and topological matter could induce transformational changes in both information and energy science.\cite{RMP 82 3045,PolitanoA,PesinD,ZhuFF} Main physical observables are Dirac states with Berry phases and high mobility due to suppressed backscattering.\cite{Dirac} This generates large linear magnetoresistance (MR), quantum Hall effect and is of high interest for electronic and optical device fabrication.\cite{RMP 81 109,Ando,XiongJ,LiangT,ShekharC,GhimireN,HuangX,GrushinAG,HillsRD}

Dirac states could be found not only at the surface of topological insulators or graphene but also in multiband bulk crystals, giving rise to magnetoresistant mobility up to $\mu$$_{MR}$$\sim$3400 cm$^{2}$/Vs, comparable to that in graphene and topological insulators.\cite{KefengSr,KefengCa,ParkSr,WangZ,WangZ2} In contrast to Na$_{3}$Bi and Cd$_{3}$As$_{2}$, ABX$_{2}$ materials feature quasi-two-dimensional (quasi-2D) electronic transport of Dirac states.\cite{KefengSr,KefengCa,LiL} Moreover, ABX$_{2}$ incorporate strongly correlated alternating Mn-Bi magnetic layers and sometimes rare earth atoms on A site.\cite{MayA,WangJK,GuoYF}  Localized magnetic moments with sufficient hybridization could provide a possible route to the Kondo effect and to transforming Dirac into a Weyl semimetal, separating the degenerate Dirac cones.\cite{MitchellA} This could result in spin-dependent band splitting and quasi-2D spin-polarized transport.

Coupling of magnons to Dirac states is indicated by Raman measurements in CaMnBi$_{2}$, where the enhanced interlayer exchange coupling drives a charge gap opening.\cite{ZhangA} In contrast, magnetic dynamics can be described by the Heisenberg model and is not influenced by Dirac fermions.\cite{RahnMC} It is of interest to establish conditions for Dirac states at the Fermi surface in  ABX$_{2}$ materials since the momentum-anisotropy of Dirac states might tailor Dirac current for applications,\cite{ParkSr,LeeG,JoY,ChoiSM,VirotF} In this study we have synthesized BaZnBi$_{2}$ single crystals that feature crystal structure identical to SrMnBi$_{2}$ and BaMnBi$_{2}$. Our first-principle calculations show presence of Dirac states near Fermi level and portions of Fermi surface with quasi-2D character, similar to other AMnBi$_{2}$ materials. However, SOC opens a larger gap in Dirac states when compared to SrMnBi$_{2}$, leaving small hole and electron pockets that dominate the electronic transport.

\begin{figure}
\centerline{\includegraphics[scale=0.4]{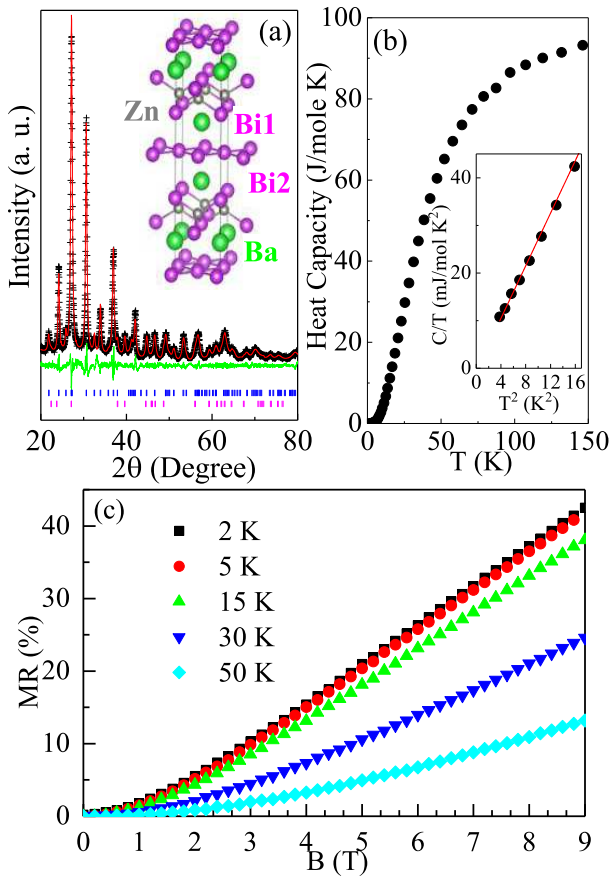}}
\caption{(Color online). (a) Powder XRD pattern and refinement results of BaZnBi$_{2}$. The data are shown by (+), fitting and difference curves are given by the red and green solid line, respectively. Ticks mark reflections
of \textit{I4/mmm} space group (upper) and residual Bi (lower). Inset in (a) shows crystal structure of BaZnBi$_{2}$. Ba atoms are denoted by green balls. Zn atoms are denoted by gray balls. Bi1 atoms denote Bi positions forming ZnBi$_{4}$ tetrahedron. Bi2 atoms denote Bi positions forming 2D square lattices. Blue lines define the unit cell. (b) Heat capacity of BaZnBi$_{2}$. (c) The magnetic field ($B$) dependence of the in-plane magnetoresistance MR at different temperatures for H//c.}
\label{magnetism}
\end{figure}

\begin{figure}
\centerline{\includegraphics[scale=0.45]{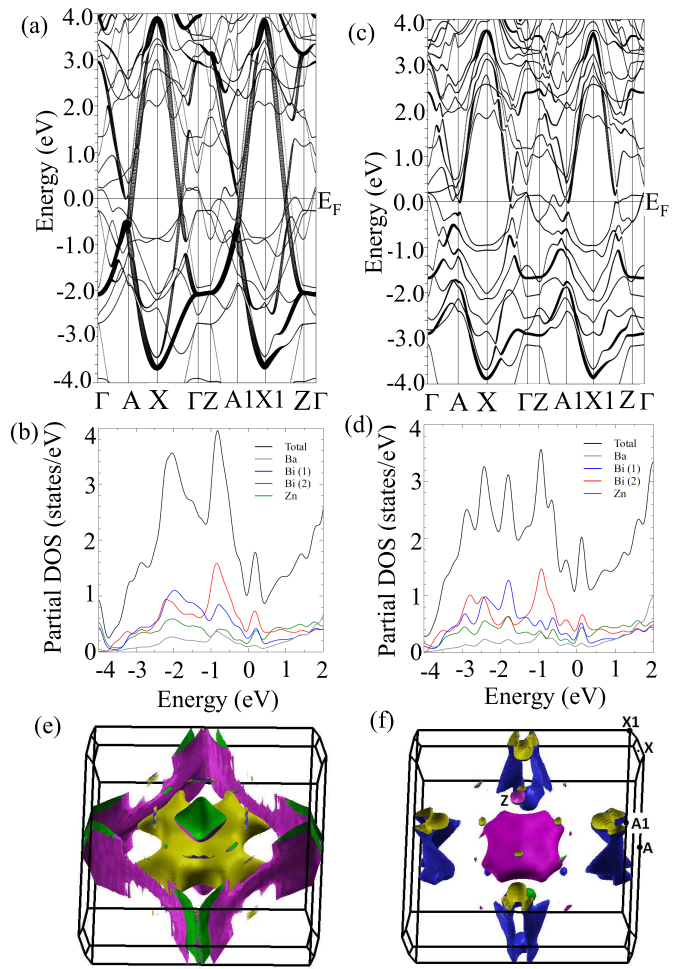}}
\caption{(Color online). The calculated band structure, total and partial density of states from Ba, Zn, Bi1 and Bi2 (a-d) of BaZnBi$_{2}$. The size of circles in (a,c) denote the 5p$_{x}$ band derived from Bi2 square lattices. The results shown are without (a, b) and with (c, d) spin-orbit coupling. Corresponding calculated Fermi surfaces are shown without (e) and with spin-orbit coupling (f).}
\label{magnetism}
\end{figure}

\begin{figure}
\centerline{\includegraphics[scale=0.3]{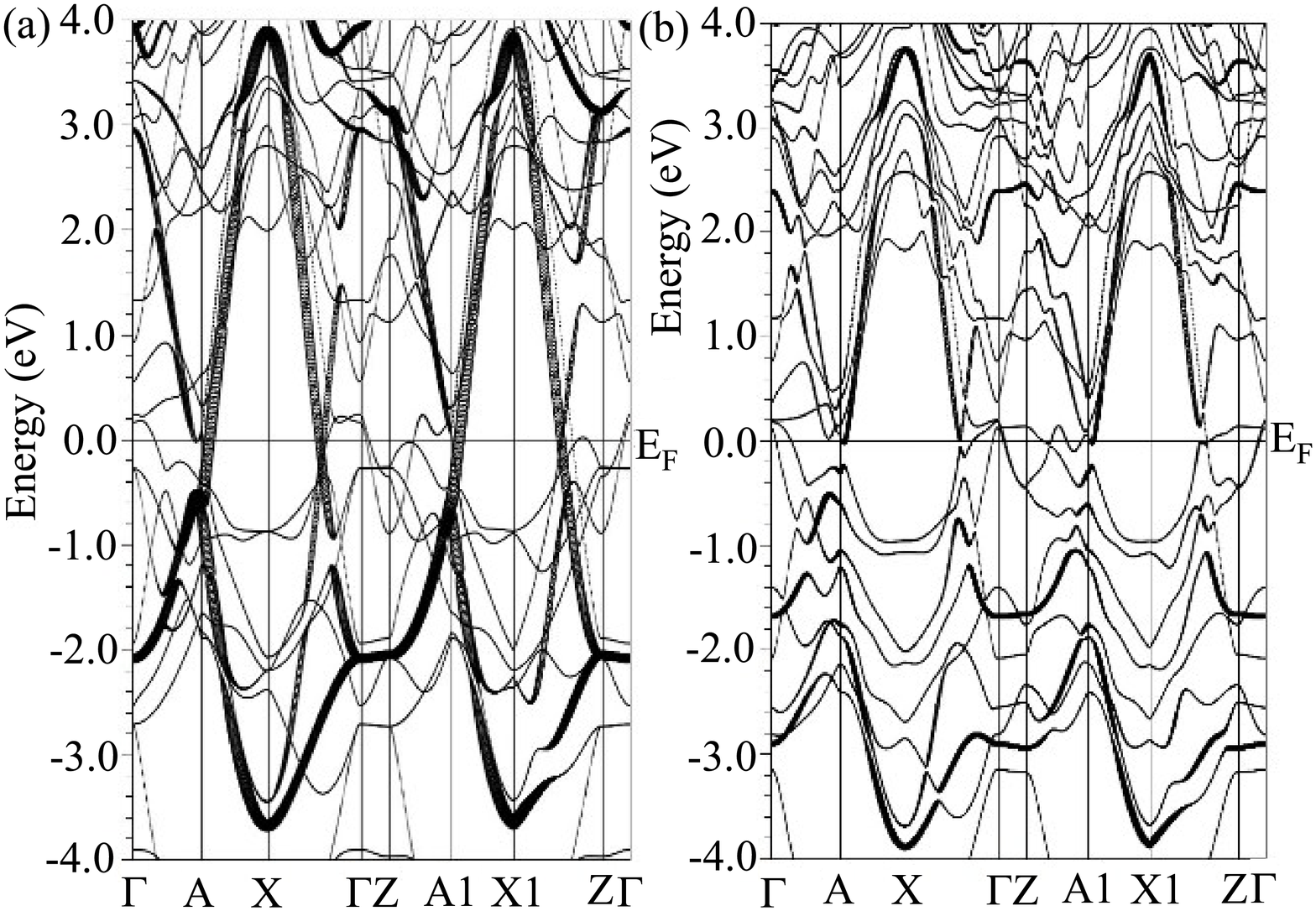}}
\caption{(Color online). (a)) Band structure calculations for hypothetical BaZnBi$_{2}$ where Ba is 5\% closer to Bi2 plane, all other parameters remaining unchanged. This leaves mass of A the same, while mimicking steric effects (lattice parameter change) due to different A size. (b) Band structure calculation for BaZnBi$_{2}$ in its true unit cell but where spin-orbit coupling is selectively switched off on the Ba atoms only. }
\label{magnetism}
\end{figure}

\section{EXPERIMENTAL DETAILS}

Single crystals of BaZnBi$_{2}$ were grown from an excess Bi flux.\cite{Fisk} Ba, Zn and Bi granules were mixed in the ratio of Ba:Zn:Bi = 1:1:10, put into an alumina crucible and then sealed in a quartz tube. The quartz tube was heated to 1073 K, held there for 6 h, then cooled to 593 K at a rate of 2.4 K/h where the excess Bi flux was decanted using a centrifuge. Shiny needle-like single crystals with typical size 5 mm $\times$ 1 mm $\times$ 1 mm were obtained. The element analysis was performed using an energy-dispersive x-ray spectroscopy (EDX) in a JEOL LSM-6500 scanning electron microscope, confirming 1:1:2 stoichiometry. X-ray diffraction (XRD) data were obtained by using Cu K$_{\alpha}$ ($\lambda = 0.15418$ nm) radiation of a Rigaku Miniflex powder diffractometer on crushed crystals. Heat capacity and magnetotransport measurement up to 9 T were conducted in a Quantum Design PPMS-9 on cleaved and polished single crystals in order to remove residual Bi droplets on the crystal surface [Fig. 1(a)]. Thin Pt wires were attached to electrical contacts made with Epotek H20E silver epoxy, producing contact resistance of about 10 $\Omega$. Sample dimensions were measured with an optical microscope Nikon SMZ-800 with 10 $\mu$m resolution. Magnetotransport in high magnetic field up to 18 T was conducted at the National High Magnetic Field Laboratory (NHMFL) in Tallahassee. Resistivity was measured using a standard four-probe configuration for both $\rho_{a}$ and $\rho_{c}$. Both measurements were conducted with 0.3 mA 16 Hz ac current excitation. All magnetotransport measurements were performed in transverse configuration with current always perpendicular to magnetic field. The de Haas van Alphen (dHvA) oscillation experiments were performed at NHMFL Tallahassee. The crystals were mounted onto miniature Seiko piezoresistive cantilevers which were installed on a rotating platform. The field direction can be changed continuously between parallel ($\theta$ = 0$^{\circ}$) and perpendicular ($\theta$ = 90$^{\circ}$) to the c-axis of the crystal. For first-principles band structure calculations, we applied the WIEN2K\cite{Blaha} implementation of the full potential linearized augmented plane wave method in the generalized gradient  approximation\cite{Perdew} of density-functional theory  with SOC treated in a second variational method. The basis size was determined by $R_\mathrm{mt}K_\mathrm{max}$ = 7 and the Brillouin zone was sampled with a regular $13\times 13 \times 13$ mesh containing 196 irreducible $k$ points to achieve energy convergence of 1 meV. The Fermi surface was plotted in a 10 000 k-point mesh.

\section{RESULTS AND DISCUSSION}

The unit cell of BaZnBi$_{2}$ crystals can be indexed in the \textit{I}4/\textit{mmm} space group by RIETICA software [Fig. 1(a)].\cite{R 1998} The lattice parameters a = b = 0.4855(2) nm and c = 2.1983(3) nm agree with the
previously reported values.\cite{BrechtelE} Hence, the BaZnBi$_{2}$ [Fig. 1(a) inset] features an identical structure, but a somewhat compressed c-axis when compared to SrMnBi$_{2}$ and BaMnBi$_{2}$. The heat capacity of
BaZnBi$_{2}$ [Fig. 1(b)] is approaching the Einstein value of 4$\cdot$3$\cdot$$R$ = 99 J mol$^{-1}$ K$^{-1}$, where $R$ is the universal gas constant 8.31 J mol$^{-1}$ K$^{-1}$. From $C/T$ as a function of $T^{2}$ at low temperature [Fig. 1(b) inset] we obtain an electronic
heat capacity $\gamma = 0.48(1)$ mJ mol$^{-1}$ K$^{-2}$ and a slope of 2.61 mJ mol$^{-1}$ K$^{-4}$. From the latter, the Debye temperature $\Theta_{D}=155(1)$ K is obtained from $\Theta_{D}=(12\pi^{4}NR/5\beta)^{1/3}$ where
$N$ is the atomic number in the chemical formula. The value of $\gamma$ is much smaller than that of the isostructural antiferromagnetic compound BaMnBi$_{2}$.\cite{LiL} Fig. 1(c) shows the magnetic field dependence of in-plane magnetoresistance MR = [$\rho$(B) - $\rho$(0)]/$\rho$(0)$(\times$100\%) at different temperatures up to 50 K. The linear unnsaturated magnetoresistance is evident at low temperatures and in high magnetic fields. Linear unsaturated magnetoresistance could be a consequence of magnetotransport  of the lowest Landau level where all components of the resistivity tensor should be
linear in temperature.\cite{Fundamentals,AbrikosovAA} Other possible mechanisms of the large linear magnetoresistance include the mobility fluctuations in a strongly inhomogeneous system, $\omega_{c}$$\tau$ $\sim$ 1 limit in thin films (where $\omega_{c}=eB/m^{*}$ is cyclotron frequency and $\tau^{-1}$ is scattering rate) and charge neutrality in a compensated semimetal.\cite{GridinVV,disorder,AlekseevPS} The former should not apply in BaZnBi$_2$ since our sample is stoichiometric crystal without doping/disorder. To get more insight into magnetotransport properties below we present results of first principle calculations.

The first-principles band structure in the absence of SOC reveals Dirac-like bands derived from the 5p$_{x}$ and 5p$_{y}$ orbitals of Bi square lattices near the X points in the Brillouin zone, which are fairly close  to the Fermi level [Fig. 2(a)]. The density of states near the Fermi level [Fig. 2(b)] is dominated by partial contribution of Bi square lattices and displays a pseudogap feature, again indicating that the Dirac points are close to the Fermi level. However, spin-orbit coupling introduces sizeable gaps into the Dirac states possibly [Fig. 2(c,d)] creating more small hole and electron pockets, while the electron states near the X points remain quasi-two-dimensional. Corresponding calculated Fermi surfaces [Fig. 2(e,f)] are quite different. This result suggests that linear MR [Fig. 1(c)] is unlikely to arise from the lowest Landau level magnetotransport.

It appears that the SOC effect on the A atoms predominates the steric effect due to different sizes of A atoms, since the same structural data were used in the calculations with and without SOC. To further distinguish the two A-site effects, we performed non-SOC calculations for BaZnBi$_{2}$ with Ba atoms moved 5\% closer to the Bi2 planes and all the other parameters remaining unchanged.  In this case, the steric effect was enhanced considerably but it has little impact on the Bi2 bands showing Dirac points [Fig. 3(a)]. In addition, we carried out calculations for the experimental structure with SOC on the Bi atoms only. We found [Fig. 3(b)] that the gapping out of the Dirac state along the $\Gamma$ - X line is nearly vanishing. This shows that the large SOC on the A-site atoms do facilitate the gapping out of the Dirac state in BaZnBi$_{2}$.

\begin{figure}
\centerline{\includegraphics[scale=0.3]{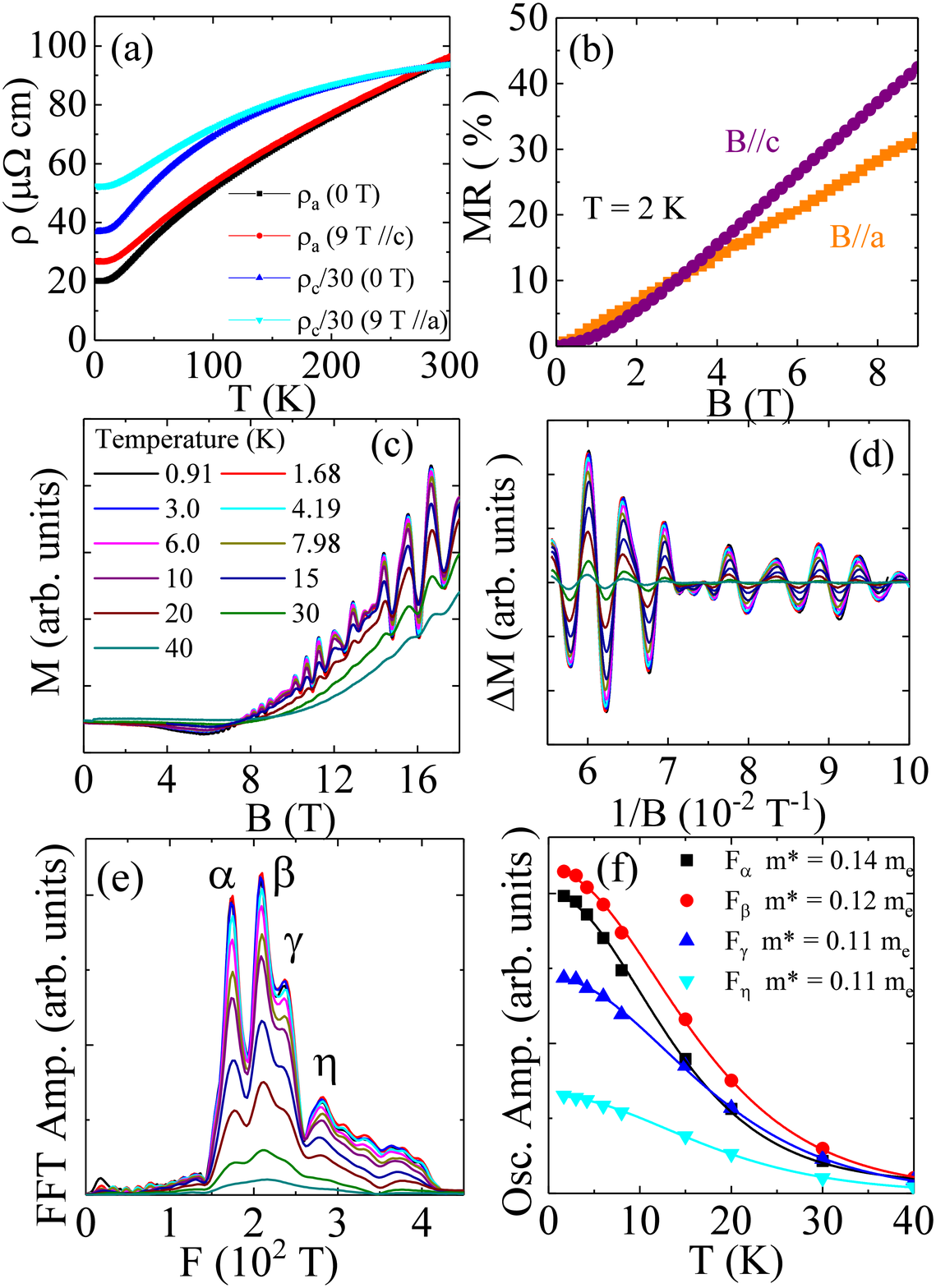}}
\caption{(Color online). (a) Temperature dependence of the in-plane resistivity $\rho_{a}(T)$ and c-axis resistivity $\rho_{c}(T)$ of the BaZnBi$_{2}$ single crystal in B = 0 and 9 T magnetic fields, respectively. Large
resistivity anisotropy indicates that the $c$-axis interlayer coupling of conductive Bi crystallographic square layers is small. (b) Field dependence of transverse magnetoresistance at 2 K for $\rho_{a}$ (filled circles) and $\rho_{c}$ (filled squares), respectively. (c) Cantilever oscillation as a function of magnetic field at different temperatures and its
oscillatory component as a function of 1/B (d). (e) FFT spectra at various temperatures, data in (c), (d), and (e) use same legend. (f) Temperature dependence of the FFT amplitudes at different frequencies, the solid line is the fitting curve using Lifshitz-Kosevich formula.}
\label{magnetism}
\end{figure}

Figure 4(a) shows the temperature dependences of in-plane ($\rho_{a}$) and out-of-plane ($\rho_c$) resistivity at 0 and 9 T for BaZnBi$_2$ single crystal. Temperature dependence of resistivity is very anisotropic($\rho_{c}$/$\rho_a$ $\sim$ 55 at 2 K and $\sim$ 30 at 300 K).  The broad hump at $\sim$ 150 K at $\rho_c$ indicates the crossover from high-T incoherent to low-T coherent conduction in a quasi-2D system.\cite{JoY} The temperature dependence of magnetoresistance (MR) for electric
current along a- and c-axes in magnetic fields up to 9 T are shown in the Fig. 4(b). The linear transverse MR is established for both current directions, however at much lower field for $\rho_c$.
The temperature-dependent cantilever signal [Fig. 4(c)] shows clear oscillations
below 40 K. The oscillatory component obtained by subtracting a smooth background is periodic in 1/B [Fig. 4(d)].
The fast Fourier transform (FFT) analysis yields four frequencies at $F_{\alpha}$ = 175 T, $F_{\beta}$ = 210 T, $F_{\gamma}$ = 237 T, and $F_{\eta}$ = 280 T, as shown in Fig.4(e).
From the Onsager relation $F$ = ($\Phi_0$/2$\pi^2$)$A_F$, where $\Phi_0$ is the flux quantum and $A_F$ is the orthogonal cross-sectional area of the Fermi surface, we estimate
$A_\alpha$ = 1.7 nm$^{-2}$, $A_\beta$ = 2.0 nm$^{-2}$, $A_\gamma$ = 2.3 nm$^{-2}$, and $A_\eta$ = 2.7 nm$^{-2}$, corresponding to about 1.0\%, 1.2\%, 1.4\%, and 1.6\% of its total Brillouin zone (BZ) in the (001) plane, respectively. These are likely to arise from the small pockets in spin-orbit induced compensated Fermi surface [Fig. 2(f)].
The cyclotron mass can be obtained from fitting the temperature dependence of the oscillation amplitude at different frequencies using Lifshitz-Kosevich formula:\cite{Shoeneberg}
$A$$\sim$[$\alpha$$m^{*}$(T/B)/$\sinh$($\alpha$ $m^{*}$T/B)] where $\alpha$ = 2$\pi$$^2$$k_{\rm B}$/e$\hbar$ $\approx$ 14.69 T/K, $m^{*}$ = $m$/$m_{e}$ is the cyclotron mass ratio ($m_{e}$ is the mass of free electron).
The fitting results are presented in Fig. 4(f), the rather small cyclotron masses corresponding to different frequencies are estimated to be m$^*_{\alpha}$ = 0.14 m$_e$, m$^*_{\beta}$ = 0.12 m$_e$, m$^*_{\gamma}$ = 0.11 m$_e$, and m$^*_{\eta}$ = 0.11 m$_e$, in agreement with heat capacity result [Fig. 1(b)].

\begin{figure}
\centerline{\includegraphics[scale=0.3]{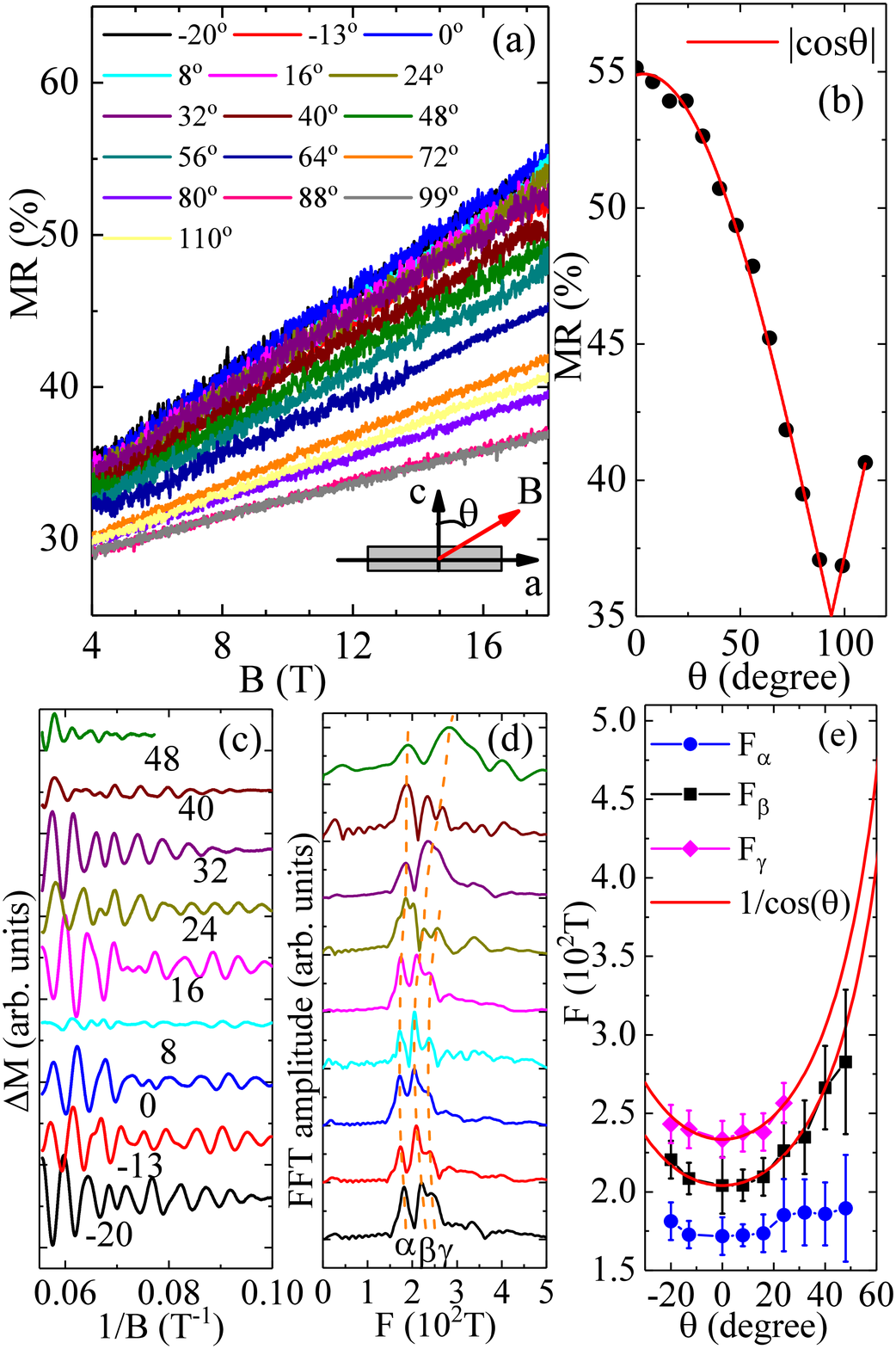}}
\caption{(Color online). (a) Angular-dependent MR for in-plane resistivity $\rho_{a}(T)$ at different tilt angles. inset cartoon shows the configuration of the measurement. (b) Tilt angle $\theta$ dependence of $\rho_{a}(T)$
at $B$ = 18 T and $T$ = 0.91 K. The red line is the fitting curve using $|$$\cos(\theta)$$|$.
dHvA oscillatory components at various title angles $\theta$ (c), and corresponding FFT spectra (d), the dashed lines are guides to the eyes. Data in (c) are using same scale and shifted vertically for clarity, while FFT spectra are normalized and shifted vertically
(c,d) Cantilever oscillation as a function of magnetic field at different tilt angles $\theta$ and corresponding FFT amplitude. (e) The angular dependence of the oscillation frequencies $F_{\alpha}$, $F_{\beta}$, and $F_{\gamma}$. $F_{\beta}$ and $F_{\gamma}$ can be fitted very well by $|$$\cos(\theta)$$|$, indicating the quasi-2D feature of $\beta$ and $\gamma$ Fermi surface.}
\label{magnetism}
\end{figure}

The response of the carriers to the applied magnetic field and the magnitude of MR is determined by the mobility in the plane orthogonal to the applied magnetic field.\cite{Fundamentals} There should be no significant
angle-dependent MR in the case of isotropic three-dimensional Fermi surface. For quasi-2D electronic systems, the 2D states respond only to the perpendicular component of the magnetic field $B|\cos(\theta)|$; hence MR oscillations
should be observed in quasi-2D conducting states.\cite{qt1,qt2,amro2} Figures 5(a,b) show the angular dependence of the $\rho_{a}$. The rotating magnetic field [Fig. 5(a) inset)] is parallel (perpendicular) to the crystallographic $c$-axis when $\theta=0^{\circ}$ ($90^{\circ}$), respectively. The $MR(\theta$) at 18 T is consistent with the $|$$\cos(\theta)$$|$ angular dependence [solid red line in Fig. 5(b)],
commonly observed in AMnBi$_{2}$ (A = Ca, Sr, Ba).\cite{KefengSr,KefengCa,LiL} This reflects the contribution of quasi-2D cyllinder-like states away from the $\Gamma$ point in the  Brillouin zone [Fig. 2(f)].

Angular-dependent quantum oscillation offfer further insight into the geometry of the Fermi surface. Cantilever oscillatory components are shown in Fig. 5(c), corresponding FFT spectra are shown in Fig. 5(d). The small oscillation amplitude at $8^{\circ}$ reflects the angle error bar in measurement in which amplitude of oscillation should be zero at strict $0^{\circ}$ and $90^{\circ}$.\cite{LiG} As shown in Fig. 5(e), $F_{\beta}$ and  $F_{\gamma}$ can be well fitted by 2D model 1/cos($\theta$), while $F_{\alpha}$ is nearly constant at different angles. $F_{\beta}$ and $F_{\gamma}$ with similar angular dependence might come from the cyllidrical states near A point, and $F_{\alpha}$ can be attributed to small  pockets near the center of BZ. On the other hand, it is difficult to trace the weak angular dependence of $F_{\eta}$ which could come from the large hole pockets around $\Gamma$ point.

It is instructive to compare mechanism of Dirac states formation with isostructural materials AMnBi$_{2}$ (A = Sr, Ba). We note that nearly-linear Dirac-like energy dispersion is evident in first-principle calculation in all materials. However, a small SOC-induced gap of about 0.05 eV appears near the Dirac point when SOC is included in SrMnBi$_{2}$.\cite{ParkSr,LeeG} SOC-induced gap in Bi2-derived states of about (0.2 - 0.3) eV is much larger in BaZnBi$_{2}$ [Fig. 2(c)], probably due to hybridization with heavier Ba near square Bi layers. This creates small pockets near $\Gamma$ point in the Brillouen zone. Together with the presence of more 3D-like Zn states not separated away from the Fermi level due to magnetic ordering, this results in the conventional magnetotransport and the absence of Dirac states in BaZnBi$_{2}$.

\section{CONCLUSIONS}

In summary, we show that  spin-orbit coupling induces gap in Dirac states, causing their removal from the Fermi surface. Resulting electronic transport is governed by cyllindrical states away from the $\Gamma$ point and small electron and hole-like pockets near Brillouen zone center. Our results suggest that the mass of the alkaline earth atoms A near Bi2 square network, in addition to magnetic order is essential for formation of Dirac cones in ABX$_{2}$ materials.

\section*{Acknowledgements}

We thank John Warren for help with SEM measurements. This work was supported by the U.S. DOE-BES, Division of Materials Science and Engineering, under Contract No. DE-SC0012704 (BNL) and the National Natural Science Foundation of
China under Grant Nos. 51671192 and 51531008 (Shenyang). Work at the National High Magnetic Field Laboratory is supported by the NSF Cooperative Agreement No. DMR-1157490, and by the state of Florida.

\end{document}